\documentclass[letterpaper,journal]{IEEEtran}
\usepackage{subfiles}
\usepackage{amsmath,amsfonts}
\usepackage{array}
\usepackage[caption=false,font=normalsize,labelfont=sf,textfont=sf]{subfig}
\usepackage{textcomp}
\usepackage{stfloats}
\usepackage{url}
\usepackage{verbatim}
\usepackage{graphicx}
\usepackage{booktabs} 
\usepackage{graphicx}   %
\usepackage{caption} 
\usepackage{listings}
\usepackage[table]{xcolor} %
\usepackage{amssymb}  %
\usepackage{pifont}   %
\usepackage{tcolorbox} %
\usepackage{multirow}

\definecolor{myblue}{HTML}{3475b3}
\definecolor{rowgrey}{RGB}{224,224,224}
\usepackage{tabularx}

\usepackage{colortbl}
\usepackage{xspace}
\usepackage{hyperref}
\usepackage{booktabs, makecell, multirow, colortbl}

\newcommand{\Ninst}{89\xspace}
\newcommand{\Nfocal}{5{,}790\xspace}

\def\BibTeX{{\rm B\kern-.05em{\sc i\kern-.025em b}\kern-.08em
    T\kern-.1667em\lower.7ex\hbox{E}\kern-.125emX}}

\newcommand{\rqone}{How effective is BreakGuard at generating valid tests?~\xspace}

\newcommand{\rqtwo}{How effective is BreakGuard at detecting breaking changes?~\xspace}

\newcommand{\rqthree}{What types of breaking failures can BreakGuard automatically detect?~\xspace}

\newcommand{\rqfour}{What is the cost of detecting breaking changes with BreakGuard?~\xspace}

\begin{document}
\raggedbottom
\title{BreakGuard: Towards Detecting Dependency Breaking Changes with LLM-Generated Tests}

\author{
  \IEEEauthorblockN{
    Rachna Raj\textsuperscript{*},
    Benoit Baudry\textsuperscript{\textdagger},
    Diego Elias Costa\textsuperscript{*}
  }
  \IEEEauthorblockA{\\
    \textsuperscript{*}Concordia University, 
    \textsuperscript{\textdagger}Universit\'e de Montr\'eal, Montr\'eal, Canada\\
    \textsuperscript{*}rachna.raj@mail.concordia.ca, 
    \quad
    \textsuperscript{\textdagger}benoit.baudry@umontreal.ca
    \quad
    \textsuperscript{*}diego.costa@concordia.ca,
  }
}
\maketitle

\begin{abstract}

Open-source libraries play an important role in software development by providing reusable features that expedite the development process. As libraries evolve, they release new versions that add features, fix bugs, or apply security patches. In this process, they may break the contract established with their clients by introducing breaking changes (BCs) that alter the runtime behavior and break client applications. Client-side test suites often fail to detect these BCs because of limited library coverage that does not exercise all library methods used in the client's codebase. 

We propose BreakGuard, an approach that generates a test suite to detect breaking changes in clients.  
BreakGuard statically extracts every client method (focal method) that invokes the target library method (call site), then generates tests per focal method. 
A test detects a BC if it passes on the pre-breaking version and fails on the breaking version. 
We evaluate our approach on 89 real-world breaking changes from the BUMP dataset, using 3 LLMs (GPT4o, Qwen3-coder-480B, GPT-OSS-120B) and three context levels: minimal, method, and class. 
Using the best-performing configuration, BreakGuard detects 30.3\% of breaking changes (27 of 89) at a mean cost of roughly \$0.90 USD per detected breaking change. 
We successfully detected BCs from different library categories (e.g., JSON libraries, logging, parsing), but we find LLM-generated tests to be more reliable for detecting crash-type breaking changes as opposed to behavioural BCs. 
\end{abstract}


\section{Introduction}

Modern software systems depend heavily on open-source libraries as dependencies, yet keeping these dependencies up to date remains a persistent maintenance challenge. When a library releases a new version, downstream clients must either update and risk breaking changes or defer the update and accumulate technical debt \cite{Venturini_2023}.
Projects consequently lag by 1.5 or more years behind available dependency versions, suggesting that the perceived risk of frequent updates outweighs the perceived cost of falling behind \cite{cox2019surviving, decan2017empirical}. 
However, deferred updates correlate with increased security vulnerabilities, missed performance improvements, and costly eventual migrations.

Several approaches help developers manage the risks associated with dependency updates. Static API compatibility analysis identifies signature-level changes, such as removed or renamed methods ~\cite{dig2006automated}, while semantic differencing targets behavioral changes but requires formal specifications that are rarely available for open-source libraries \cite{Glock_2024}. These approaches can identify potential incompatibilities, but they do not establish whether a change actually affects a particular client. Determining whether a client is affected requires exercising the interactions between the client and the updated library.
Existing test suites provide limited evidence about these interactions. Client-side tests typically focus on their own business logic and often mock external dependencies, leaving the actual library interactions unexercised \cite{chu2026llmtestsurvey}. Conversely, library test suites evaluate API correctness but do not capture the diverse conditions under which different clients use those APIs \cite{raj2025supportingopensourcelibrary}. Consequently, neither side necessarily tests every library API method a client uses; updates may break interactions involving a particular client. Such breaking changes may therefore remain undetected until production.

This gap motivates the need for automatically generating tests that exercise client-specific library usage and evaluate it across library versions. Large Language Models (LLMs) have demonstrated strong capabilities in software development and maintenance, including unit test creation \cite{yuan2023chatunitest, schafer2023empirical}, test oracle generation ~\cite{bodicoat2025understanding}, and assertion synthesis ~\cite{roychowdhury2025staticprogramanalysisguided}. Research also proposes techniques combining program analysis with LLM prompting to improve coverage and precision \cite{Nan2025Test}. However, existing work primarily generates tests for internal program logic rather than for evaluating whether dependency updates preserve client-library interactions across versions. To the best of our knowledge, no prior study has evaluated the effectiveness of LLM-generated tests for detecting breaking changes caused by library version updates.

To address this gap, we conduct an empirical study of breaking-change detection using LLM-based test generation and propose BreakGuard, an automated approach that generates tests that exercise every client-library interaction and execute them across library versions. Given a client migrating from one library version to another, BreakGuard first identifies the library APIs invoked by the client code. It then generates migration test suites using prompts that vary in the amount of API and usage information provided to the LLM, and executes the generated tests against both the pre-breaking and breaking library versions. A difference in test outcomes provides evidence that the library update introduces a breaking change.

We evaluate BreakGuard on real-world breaking changes from the BUMP dataset~\cite{reyes2024bump}. 
BUMP provides a reproducible benchmark consisting of breaking-update instances of a client-library pair migrating from one library version to another, with tests that capture breaking dependency updates in Java, packaged as a Docker image.  Our study covers 89 breaking-update instances across 25 OSS libraries and 31 client projects. We compare three context variants across three LLMs (GPT-4o~\cite{openai2024gpt4ocard}, Qwen3-Coder~\cite{qwen3coder2025}, GPT-OSS~\cite{openai2025gptoss120bgptoss20bmodel}). Our study addresses the following research questions:

\textbf{RQ1: \rqone} 
We measure the compilation and execution success of generated tests on the pre-breaking version.
    
\textbf{RQ2: \rqtwo} 
We measure the breaking change detection rate by running all valid tests from the pre-breaking version. Tests that pass on the pre-breaking but fail on the breaking version successfully detect the breaking change.

\textbf{RQ3: \rqthree}
We characterize the failure mechanisms of tests that detected breaking changes and examine whether generated tests expose crash-type or behavioral breaking changes.

\textbf{RQ4: \rqfour} We measure the token and monetary cost of test generation to detect breaking changes across models and context variants and examine how cost scales with the number of library call sites.

Our results show that richer context (full class code rather than method signatures alone) provides the best trade-off between detection effectiveness and cost, detecting 30.3\% of breaking changes at a median cost of approximately \$0.09 per breaking-update instance. Generated tests predominantly detect crash-type breaking changes, while behavioral changes remain largely undetected. The detected breaking changes are not confined to a single library type. BreakGuard surfaces breaking changes in JSON libraries, logging, parsing, application-framework, and utility libraries, though the specific categories covered vary by model and context.
Overall, this paper makes the following contributions:

\begin{enumerate}

\item An approach for generating test suites from an LLM for a library dependency used by clients to detect breaking changes in client code. 

\item An empirical study on 89 real-world breaking changes, examining three LLMs and three context levels.

\item A characterization of the types of breaking changes detected by LLM-generated tests, providing insights into its effectiveness and limitations.

\item A publicly available prototype of our approach \textbf{BreakGuard} that applies to any client project in a pull-request workflow. Both the pipeline and code are available on GitHub.

\end{enumerate}

\section{Motivation}
\label{sec:background}

\begin{figure}
    \centering
    \includegraphics[width=.5\textwidth]{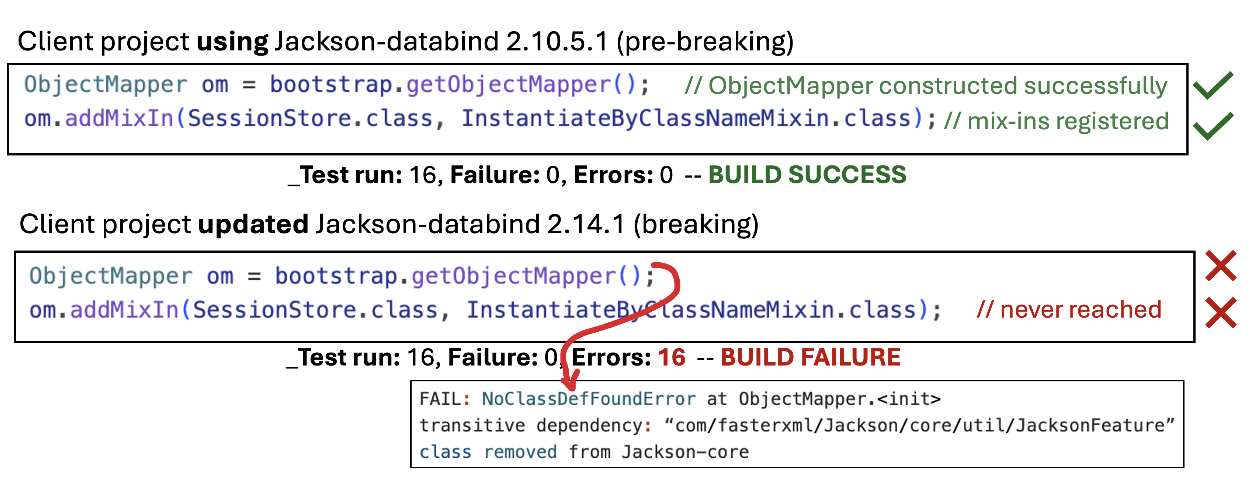}
    \caption{A breaking change in jackson-databind when migrating from version 2.10.5.1 to 2.14.1 }
    \label{fig:BC_example}
\end{figure}

\begin{figure}
    \centering
    \includegraphics[width=.5\textwidth]{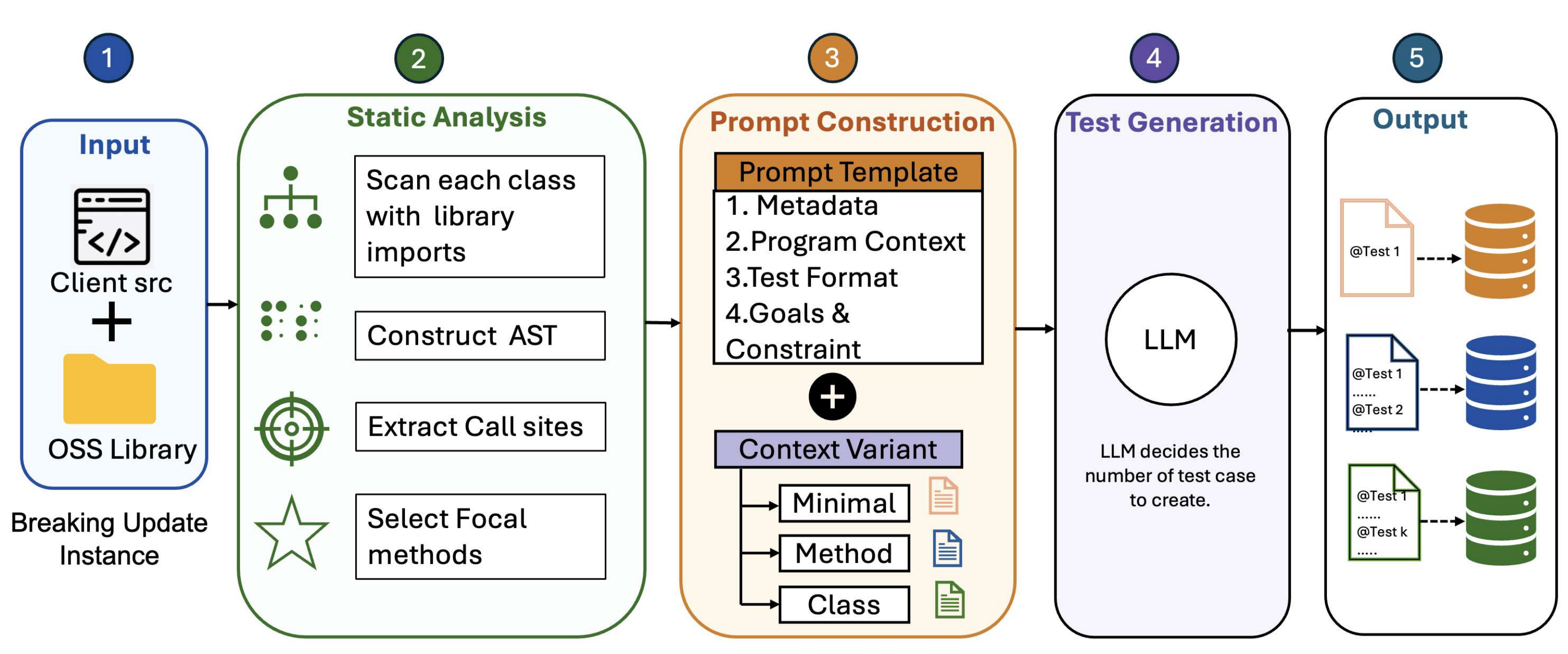}
    \caption{Pipeline Overview of BreakGuard}
    \label{fig:Overview}
\end{figure}

Breaking changes can only be detected by client tests when those tests exercise the affected API call sites. However, a client typically uses only a subset of a library's API, and its test suite may provide little or no coverage for some of these API call sites \cite{raj2025supportingopensourcelibrary, harrand2022api}. As a result, breaking changes in untested API call sites can remain undetected until they cause failures after an upgrade.

Consider a real-world example from the BUMP dataset, shown in Figure~\ref{fig:BC_example}. The client project \textit{dropwizard-pac4j} uses Jackson-databind to configure how objects are serialized. In particular, it calls \texttt{ObjectMapper.findMixIn(...)} to register a \emph{mix-in}, a helper class that tells Jackson how to serialize a target class without modifying that class itself. When Jackson-databind is upgraded from version 2.10.5.1 to 2.14.1, the client code and the API signatures remain unchanged. However, the new Jackson-databind version references \texttt{JacksonFeature}, a class that was removed from its transitive dependency, Jackson-core. As a result, constructing an \texttt{ObjectMapper} throws a \texttt{NoClassDefFoundError} at runtime, before the mix-in can even be registered.
 
 Here, the client's existing test suite happened to exercise this construction path, so the regression was detected: the same 16 tests that pass on the pre-breaking version all fail on the breaking version with the same error. However, such coverage cannot be assumed for other API call sites. A client may contain hundreds of API call sites, while its tests exercise only some of them \cite{RegressionTesting}. Manually writing tests for all such API call sites before a dependency upgrade is impractical. This is where \textbf{BreakGuard} can help: rather than relying only on existing client tests, it automatically generates tests for API call sites in the client code and executes them against both the old and new dependency versions to proactively detect breaking changes before they reach production.

\section{BreakGuard: Automatic Generation of Breaking Change Detectors} 
\label{sec:breakguardDesign}

Our goal is to investigate the effectiveness of LLM-generated tests for detecting breaking changes in real-world client projects. For that purpose, we develop BreakGuard, an approach that generates unit tests to detect breaking changes when a client project migrates to a new library version (Figure~\ref{fig:Overview}). 
BreakGuard receives the client project source code as input and the target library to be migrated. 
The core idea behind BreakGuard is to test all the usages of a library to identify breaking changes, 
as a client project may invoke different library methods, and any of them can behave differently after an upgrade.

\begin{itemize}
    \item \textbf{Input:} the client project's source code and the target library's compiled bytecode.
    \item \textbf{Output:} a migration test suite, a set of tests that exercise the client's library usage.
\end{itemize}

BreakGuard organizes test generation around the concept of focal methods. 
A \textbf{focal method} is a client method that contains at least one statement that invokes a method of the target library, which we refer to as a~\textbf{library call site}. 
BreakGuard scans the client project for every library call site, groups them by their enclosing focal method, which becomes the unit of test generation. 
BreakGuard generates one test file per focal method, but we do not enforce the number of test cases generated.   
Therefore, each test file contains one or more test cases. 
Together, all tests generated for the client project form its \textbf{migration test suite}. 

BreakGuard is designed to exercise every library call site used by the client. An optimized alternative for addressing this problem is to ask the LLM to generate tests only for call sites that have changed between the two library versions (a diff-based approach). 
We do not do this for two reasons. First, diff-based generation is prone to missing behavioral breaking changes. Many library updates modify internal logic without altering the visible API signature \cite{jayasuriya2025extended, brito2018and}, meaning surface-level diffs would fail to trigger test generation for those affected sites.
Second, BreakGuard is meant to build a migration test suite before any upgrade is planned. Because client projects often lack tests specifically for their library call sites \cite{hejderup2022trust, client-lib-compatibility-testing}, BreakGuard solves this gap by exercising the client's library API usage to generate a standalone suite. This means the resulting tests are not tied to a single version change, but can be reused for any future library upgrades. Section~\ref{sec:methodology} describes how we apply it to the BUMP benchmark.

\subsection{Static Analysis: Extracting Focal Methods and API Call-site}
\label{sec:static-analysis}

Given a client project and a target library to migrate, BreakGuard's static analysis aims to extract every focal method with one or more call sites in three steps

\textbf{1) Import filtering:} We scan every source file in the client project for import statements matching the library's package prefix and keep only the files that import the library at all. This narrows the set of files that require full AST analysis. Note that in our evaluation on the BUMP dataset, we also excluded all test files from the analysis, as the tests were shown to detect the breaking change.

\textbf{2) Call site extraction:} We build an AST for each filtered client source file using Spoon~\cite{spoon2015}, an AST-based Java source code analysis framework. 
For every method in the client code, we extract two kinds of information about the target library.

\begin{itemize}
    \item \textbf{Library method invocations:} each call site, with the method name, argument types and expressions, return type, and signature of the invoked library method. 
    \item \textbf{Type references:} references to library types in variable declarations, parameters, return types, and generic arguments.
\end{itemize}

\textbf{3) Focal method grouping:} We group call sites by their enclosing focal method, preserving the order in which the client invokes the library. A focal method becomes the target for test generation regardless of visibility (public or private), as the generated test are meant to be exercised as part of the client test suite.

\subsection{Prompt Construction}
\label{sec:prompt-construction}

\begin{figure}
    \centering
    \includegraphics[width=.5\textwidth]{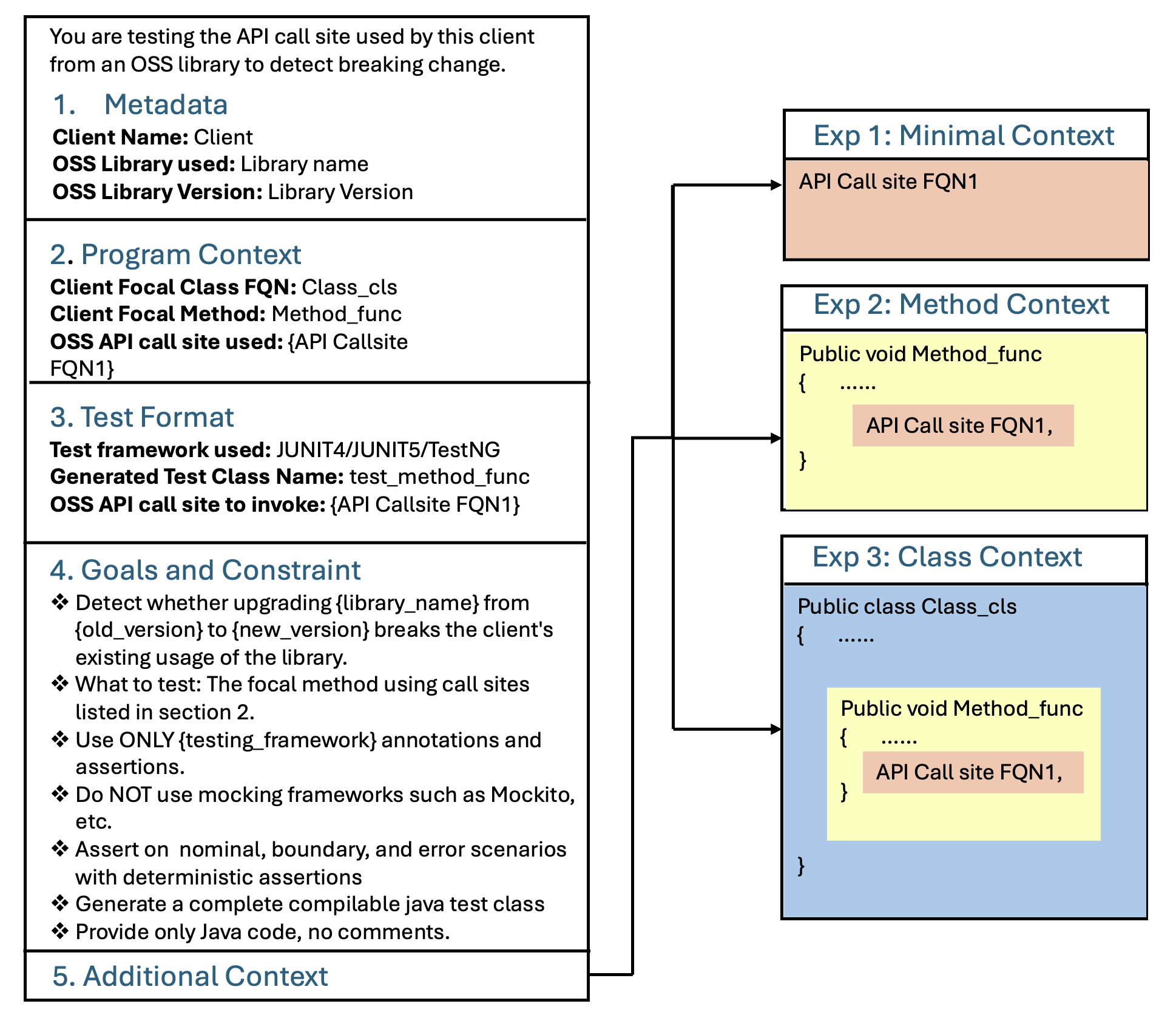}
    \caption{Prompt Format of BreakGuard}
    \label{fig:promptdesign}
\end{figure}

For each focal method, we construct a prompt and submit it to an LLM to generate a test file. The prompt design follows Nan et al.~\cite{Nan2025Test} for intention-guided unit test generation and comprises five sections. Figure \ref{fig:promptdesign} presents our prompt design. 

\begin{itemize}
    \item \textbf{Metadata} gives the client project name, the library name, and the pre-breaking and breaking version details.

    \item  \textbf{Program Context} provides the focal method signature and its call sites, so the LLM can see how the client actually uses the library.

    \item \textbf{Test Format} specifies the test framework (JUnit~4, JUnit~5, or TestNG), the test class name, and the annotation structure. 

    \item \textbf{Goal \& Constraints} states what to generate and what to avoid. We ask for a complete, compilable test class with deterministic assertions covering nominal, boundary, and error scenarios. We do not fix how many test cases to write; the LLM decides, and each scenario becomes its own test case in the same test file. We prohibit mocking frameworks such as Mockito because a mocked test replaces the real library with a fake and cannot reveal version-specific changes. We require the LLM to avoid unused imports and explanatory text, so generated tests run with less manual editing.
    
\end{itemize}

The fifth section provides \textbf{Additional Context}, where we explore three variants to assess how much code context the LLM needs to generate effective tests. 

\begin{itemize}
    \item \textbf{Minimal} provides only the focal and call-site method signatures. This is the lowest cost and most focused variant, keeping the prompt short.

    \item \textbf{Method} adds the focal method's full source code. This lets the LLM see the surrounding statements, local variable types, and control flow around each call site, so it can reconstruct the objects and arguments the client actually passes rather than inferring them from the signature alone. 

    \item \textbf{Class} adds richer context by providing the entire source code of the class, including fields, imports, helper methods, and construct logic. This gives the LLM the setup code and details, which are often required to instantiate the focal object that can reach the library call in a compilable test. 
\end{itemize}

\subsection{Test Suite Generation}
\label{sec:test-suite-generation}

Once we construct a prompt for every focal method, we submit the prompt to an LLM and receive a candidate test file in response. 
We verify that the response contains only valid Java code, with no explanatory text, and strips any residual formatting artifacts. 
Repeating this step for every focal method identified in a client project, BreakGuard produces the migration test suite for that breaking update instance (client-library pair). 
Our approach is not tied to any specific LLM and serves as a test-generation backend.

\begin{figure*}
    \centering
    \includegraphics[width=0.9\textwidth]{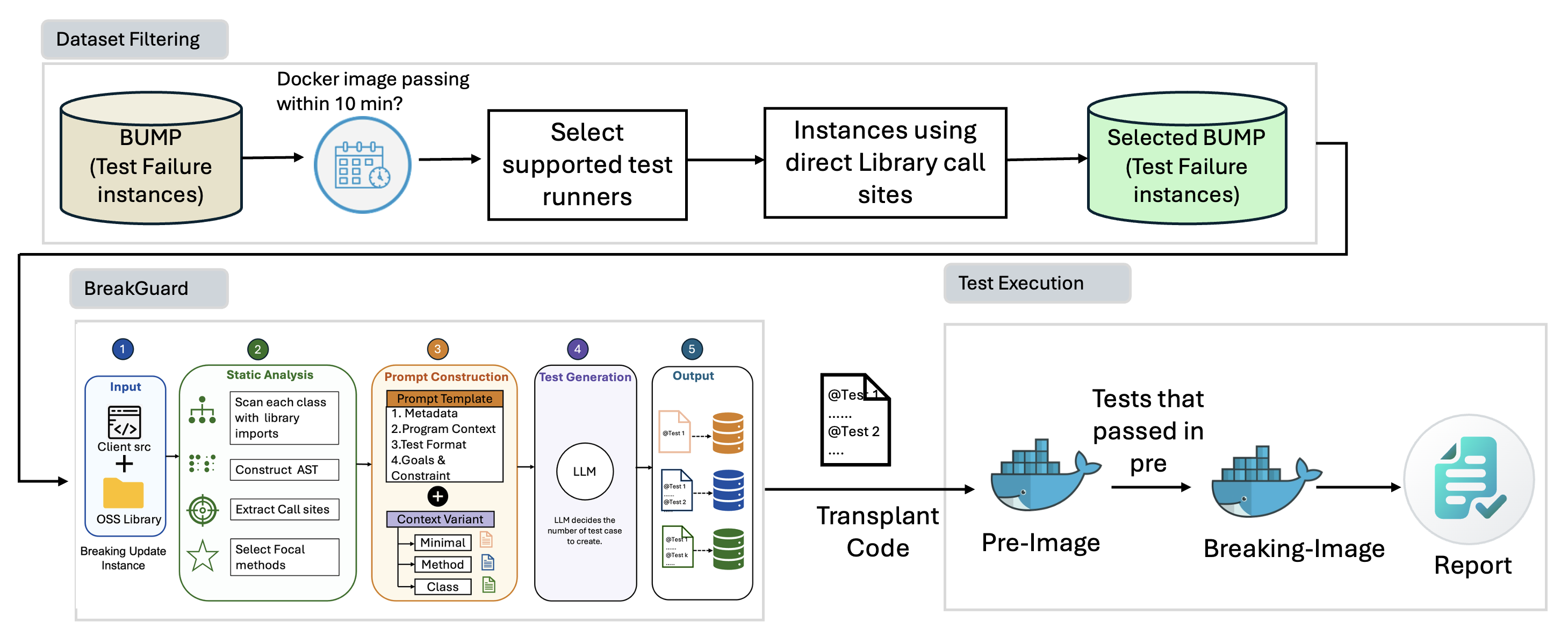}
    \caption{Overview diagram of applying BreakGuard on BUMP.}
    \label{fig:bumpWithBreakGuard}
\end{figure*}

\section{Experiment Setup}
\label{sec:methodology}

We evaluate the effectiveness of BreakGuard on real-world breaking changes using the BUMP benchmark~\cite{reyes2024bump}. 
Figure~\ref{fig:bumpWithBreakGuard} gives an overview of this experimental setup, fully detailed in the remainder of this section.

\begin{table}
\centering
\caption{Dataset Filtering Steps.}
\label{tab:bump_categories}
\begin{tabular}{lrr}
\toprule
\textbf{Filtering Step} & \textbf{Remaining} & \textbf{Removed} \\
\midrule
BUMP breaking changes (total)     & 571 & ---  \\
Test failure instances only        & 188 & 383  \\
Docker timeout filtering (10 min)  & 179 &   9  \\
Test runner extraction             & 164 &  15  \\
Static analysis                    & 104 &  60  \\
Call sites in test files only      &  89 &  15  \\
\midrule
\textbf{Final dataset} & \multicolumn{2}{c}{\textbf{89 }} \\
\bottomrule
\end{tabular}
\end{table}

\begin{table}
\centering
\caption{Library categories in the retained dataset}

\label{tab:library-scope}
\footnotesize
\begin{tabular}{lll}
\toprule
\textbf{Maven Category} & \textbf{Example Libraries} & \textbf{Instance} \\
\midrule

JSON Libraries & jackson-databind, jackson-core & 27 \\
Logging API & slf4j, log4j, google extensions & 26 \\
Application framework & spring-tx, spring-webmvc & 9 \\
Web server / Servlet & jetty-server & 6 \\
Parser generation & antlr4-runtime & 6 \\
Other Utilities & pdfbox, sshd-common, h2 & 6 \\
Mocking library & mockito-core & 4 \\
Maven Plugins & plexus-utils, versions-maven-plugin & 3 \\
HTTP Clients & Apache hc, client-extensions & 2\\
\bottomrule
\end{tabular}
\end{table}

\subsection{\textbf{Dataset Filtering}}
\label{sec:dataset}

To evaluate the effectiveness of BreakGuard, we opt to select a dataset of real-world breaking dependency updates with known ground truth.
To the best of our knowledge, BUMP~\cite{reyes2024bump} is the only reproducible benchmark of real-world breaking dependency updates in Java, containing 571 breaking updates from 153 projects.
Each breaking update instance in BUMP represents a real-world breaking dependency update in a client project, packaged as a pair of Docker images to allow reproducible execution. For ease in writing and understanding, we will call the \textbf{breaking-update instance} as \textbf{BUMP instance} from now onwards.

We restrict our BUMP selection to instances that satisfy the following criteria, summarised in Table~\ref{tab:bump_categories}. Note that while some criteria are determined during static analysis (Section~\ref{sec:static-analysis}), we present them all here together for clarity.

\begin{itemize}
    \item \textbf{Test failure instances.} We focus exclusively on the 188 BUMP instances categorized as test failures. Unlike compilation or resolution errors, these represent breaking changes that manifest as regressions during test execution: making them the ideal ground truth for evaluating whether LLM-generated tests can detect breaking changes based on client-side API usage patterns.

     \item \textbf{Executable within resource constraints.} We exclude 9 BUMP instances where Docker image execution did not complete within 10 minutes, as these resource-intensive builds make test execution infeasible at scale, leaving 179 BUMP instances.

    \item \textbf{Supported test runner.} To generate syntactically correct tests, we target specific test frameworks. We focus on client projects using JUnit~4, JUnit~5, or TestNG, as these are the most widely used Java test frameworks~\cite{lops2025llmsautomatedunittest}. BUMP Instances using other or unidentifiable frameworks are excluded, removing 15 instances and leaving 164.

    \item \textbf{Direct library API call-sites in client's code.} 
    We perform static analysis on each BUMP instance to extract all OSS library API call-sites from client code (Section~\ref{sec:static-analysis}). We exclude 60 BUMP instances where we could not directly find API usage in the client code. One scenario is when the client accesses the library only via transitive dependencies or through reflection and dynamic class loading. 
    We further exclude 15 BUMP instances where API call-sites are detected exclusively in the client's existing test files, as using them as generation context would not reflect our intended scenario of generating tests from production code usage alone.

\end{itemize}

After applying all criteria, \textbf{89 breaking-update instances} remain, each with a feasible Docker environment, a supported test runner, and detectable direct library API call-sites in production code, spanning  31 client projects, which use 25 different OSS libraries.
Table~\ref{tab:library-scope} summarizes the library categories represented in the retained BUMP instances. It presents the library categories used in the client project. We extract category information for each library from Maven Central \cite{mvnrepository_opensource}. JSON Libraries (jackson-databind, jackson-core) and logging APIs (slf4j-api, log4j-api, log4j-core) together account for the majority of the 89 retained instances.

\subsection{LLM Selection for Experiments}

We selected three large language models (LLMs) based on their state-of-the-art performance on coding tasks, drawing on established evaluation frameworks for LLMs in code generation \cite{silva2025repairbench, reyes2026byamfixingbreakingdependency}. Table \ref{tab:llm_comparison} summarizes the key characteristics of the selected models. Our selection covers different model families and access modes:

\begin{itemize}
    \item \textbf{GPT-4o} (OpenAI), because it consistently ranks among the top-performing closed-source models on coding benchmarks at the time of writing, making it a strong commercial baseline for test generation.
    \item \textbf{Qwen3-Coder (480B)} (Alibaba), because it is the leading open-source code-specialized model at large scale reported by \cite{reyes2026byamfixingbreakingdependency} and during the time of this experiment. The selection was to assess whether open-source models can match closed-source models in library migration test generation.
    \item \textbf{GPT-OSS (120B)} (OpenAI), because it offers a cost-effective open-source alternative while maintaining strong code generation capabilities.
\end{itemize}

Table~\ref{tab:llm_comparison} summarizes the parameter configuration used for all three models. To ensure a fair 
comparison, we explicitly set \textit{temperature=0} and \textit{top\_p=1.0} for all models to steer the outputs towards determinism \cite{silva2025repairbench}. We left Parameters such as \textit{top\_k} and \textit{repetition\_penalty} at their respective defaults.

\begin{figure}
    \centering
    \includegraphics[width=.5\textwidth,height=2in,keepaspectratio]{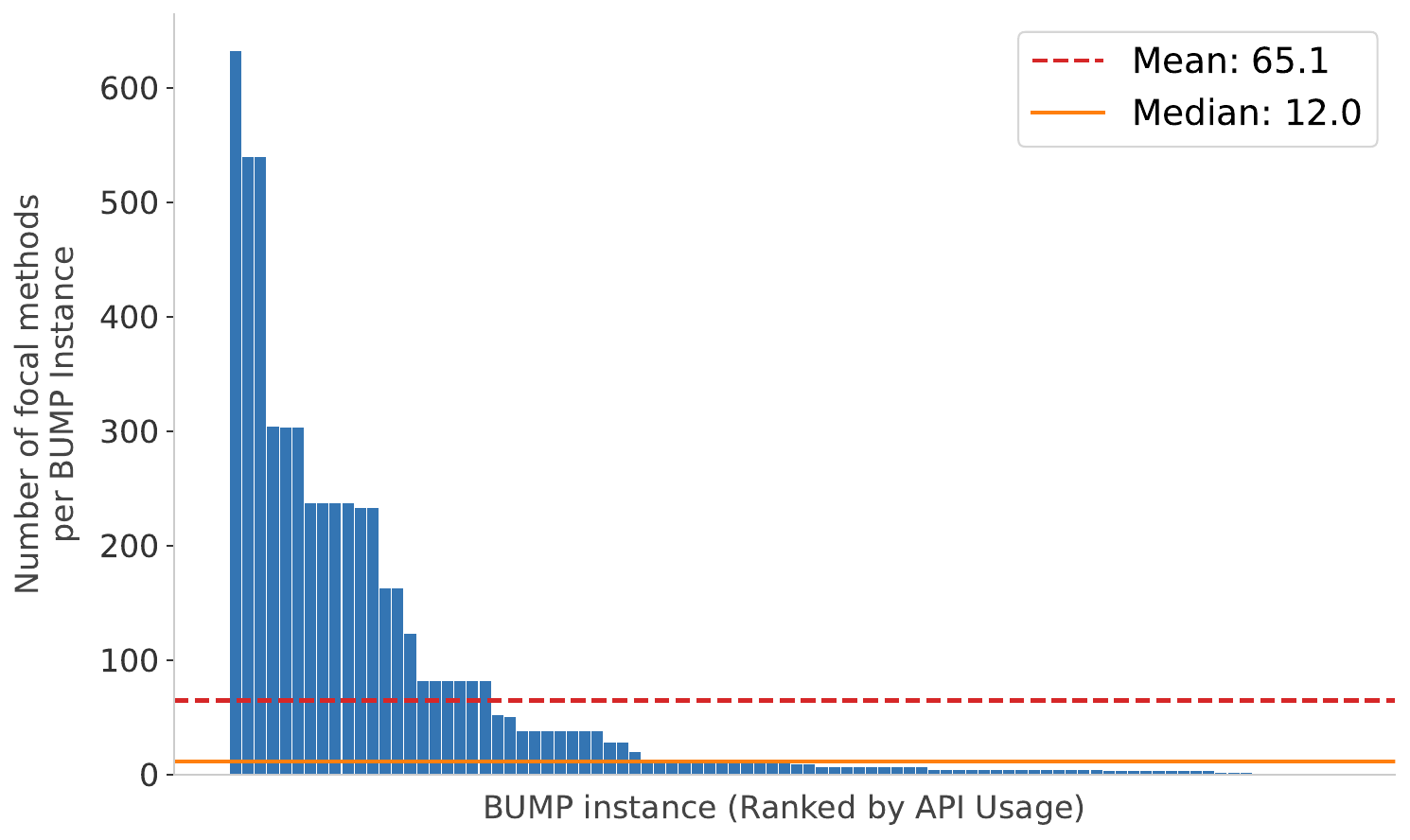}
    \caption{Ranked bar chart of focal methods per BUMP instance. Each bar represents an individual BUMP instance, ranked from the highest number of valid focal methods (left) to the lowest (right). The dashed and solid horizontal lines denote the mean and median values, respectively.}
   
\label{fig:APICALLSIGHTSTATS}
\end{figure}

\begin{table}
\centering
\caption{LLMs used in our experiments. All models were configured with
\textit{temperature}=0 and \textit{top\_p}=1.0.}
\label{tab:llm_comparison}
\begin{tabular}{llll}
\toprule
\textbf{Feature} & \textbf{GPT-4o} & \textbf{Qwen3-coder} & \textbf{GPT-OSS} \\
\midrule
Provider         & OpenAI      & Alibaba       & OpenAI \\
Parameters       & Undisclosed & 480B          & 120B \\
Inference        & OpenAI API  & Ollama Cloud  & Ollama Cloud \\
Open Source      & No          & Yes           & Yes \\
Input Tokens     & 128K        & 131K          & 128K \\
Output Tokens    & 16K         & 131K          & 16K \\
Knowledge Cutoff & Oct 2023    & Undisclosed   & Undisclosed \\
\bottomrule
\end{tabular}
\end{table}

\subsection{Applying BreakGuard to the Selected BUMP Instances}
\label{sec:applying-breakguard}

After BreakGuard generates a migration test suite for each BUMP instance, we need to execute the test suite to assess its ability to detect breaking changes.   
First, BreakGuard runs static analysis (Section~\ref{sec:static-analysis}) on each of the 89 selected BUMP instances one by one at the pre-breaking version of the client's code. 
Across all 89 BUMP instances, this process identifies \textbf{5,790 distinct focal methods}, each exercising one or more call sites to the target library. 
Note that the number of focal methods exercising call sites varies across BUMP instances, as some client projects make heavier use of the target library than others. Figure~\ref{fig:APICALLSIGHTSTATS} shows how the 5,790 focal methods ranked from highest to lowest. Most BUMP instances contain a small number of focal methods, while a few instances make heavy use of the target library: 3 of the 89 BUMP instances contribute more than 500 focal methods each. 

Second, for each focal method, BreakGuard constructs a prompt and generates one test file, yielding 5,790 test files in total. 
We repeat this generation step once per LLM-context variant pair, so each selected model produces its own 5,790 test files under Minimal context, another 5,790 under Method context, and another 5,790 under Class context. 
In total, our experiment setup produces nine migration test suites, one per model-variant combination, each containing 5,790 test files.

\subsection{\textbf{Test Execution}}

To evaluate the quality of the generated migration test suites, we run the entire suite against both the pre-breaking and breaking versions of the BUMP instance.
The pre-breaking version refers to the client project immediately before the breaking update, which depends on the original library version. 
Conversely, the breaking version refers to the client project version after the breaking update, where the client has migrated to a new library version. 

While executing the migrated test suites, we identified three challenges to ensure the soundness of our evaluation:  
\begin{itemize}
    \item \textbf{Test interference:} our experiment should ensure that each test run's success or failure does not depend on the execution of other tests. We address this problem by running each generated test in isolation. We mount only one generated test file at a time to record its success and hide the project's original tests. Finally, we run \texttt{mvn test -Dtest= \{testFile}\} on the mounted test file to ensure that no state, classpath, or side effects are shared between runs.

    \item \textbf{Reproducibility:} our experiment should ensure the same environment (e.g., classpath, dependencies) across all tests. 
    The BUMP benchmark provides pre-built Docker images for each instance in two versions, tagged \texttt{-pre} and \texttt{-breaking}, with a complete, resolved build environment.
    We use the provided Docker images, and every run uses the same resolved classpath and dependency resolution, so results do not vary between runs. 
    We achieve this without modifying the client project's own dependency declarations.

\item \textbf{Observability:} Our experiment must distinguish between tests that fail due to a real breaking update and those that fail for other reasons, such as a misconfigured environment. To prevent environment issues from being mistaken for test failures, we perform a two-level canary validation. We first verify that the client's Docker image builds and runs successfully within a 10-minute timeout. We set the timeout to make overall execution time feasible across 5,790 test files. Next, we run a minimal dummy test (e.g., asserting "Hello World") to confirm that the client's test framework is correctly wired up and executable.
Only instances that pass both canary checks are used to evaluate our generated tests.
    
\end{itemize}

\begin{figure}
    \centering
    \includegraphics[width=.4\textwidth]{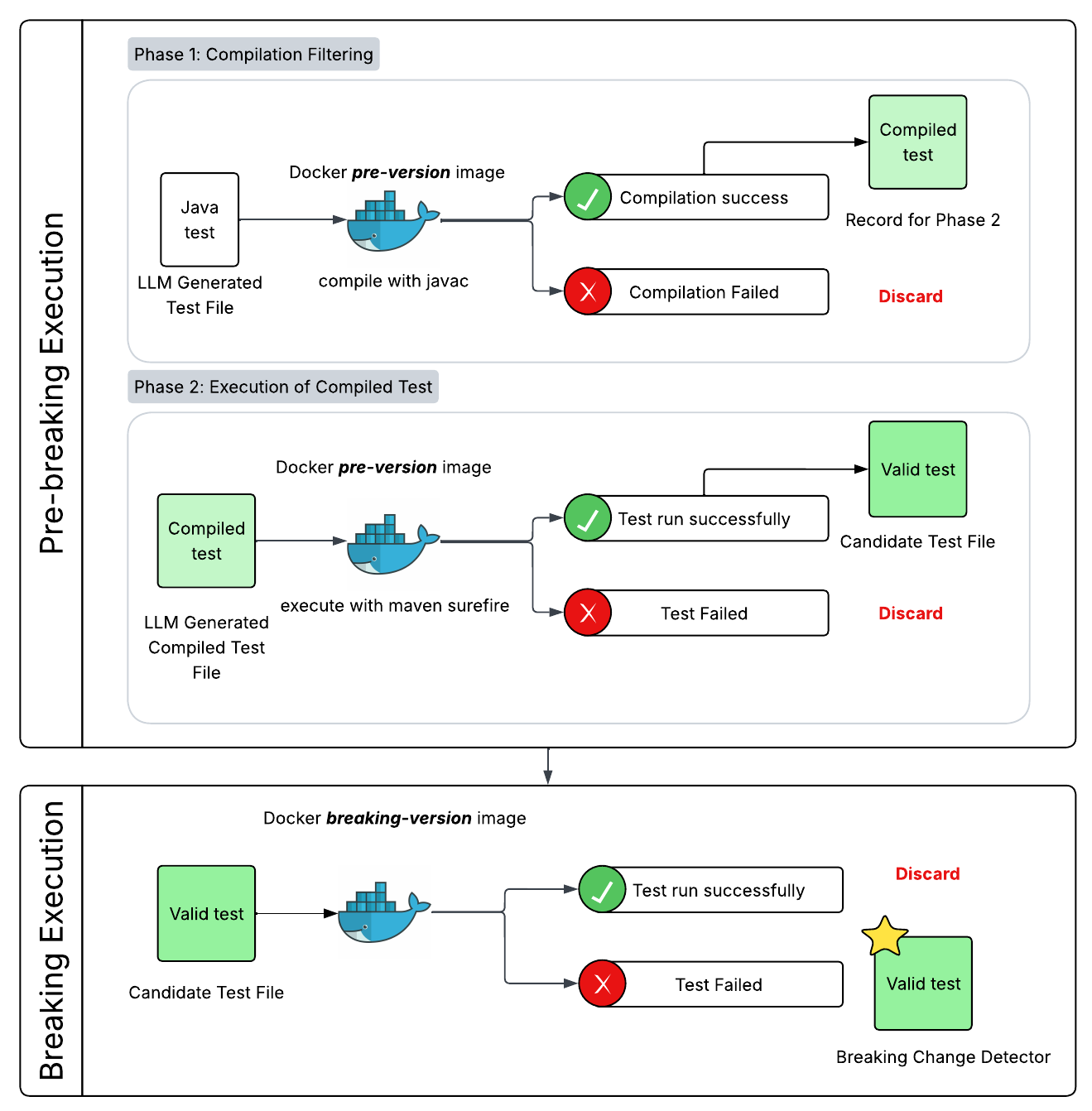}
    \caption{Overview of Two-Phase Docker Execution Pipeline}
    \label{fig:twophaseexecution}
\end{figure}

We execute all tests in two phases, separating compilation from execution to optimize time and resource usage (Fig.~\ref{fig:twophaseexecution}). Phase 1 compiles every generated test file against the project's classpath and records which tests compile successfully. Phase~2 executes the compiled tests and records which ones pass. This forms a \textbf{valid test} for evaluating the breaking version. We run this two-phase pipeline twice per instance: first on the pre-breaking version, to establish the set of valid tests; then, using only that valid set, on the breaking version, to check for detections.

\section{Results}
\label{sec:results}

This section presents the results and insights from our findings on the three proposed RQs.

\subsection{RQ1: \rqone}
\label{sec:rq1_results}

\begin{table*}
\centering
\footnotesize
\caption{Validity of BreakGuard-generated tests on the pre-breaking version, per model and
context variant. A focal method is considered covered if at least one test compiles and passes on the pre-breaking version. Similarly, a BUMP instance is covered if at least one of its tests is valid. The three highest values in each valid test coverage column are \colorbox{gray!20}{shaded}.}
\label{tab:rq1_quality}
\begin{tabular}{@{}llrrrr@{}}
\toprule
& & \multicolumn{2}{c}{\textbf{Valid test coverage}}
  & \multicolumn{2}{c}{\textbf{Reason for failure}} \\
\cmidrule(lr){3-4} \cmidrule(l){5-6}
\textbf{Model} & \textbf{Context}
  & \textbf{Focal methods} & \textbf{BUMP instances}
  & \textbf{Compilation} & \textbf{Test failure} \\
& & (of \Nfocal) & (of \Ninst) & \multicolumn{2}{c}{(\% of \Nfocal)} \\
\midrule
\multirow{3}{*}{GPT-4o}
    & Minimal &  418 (7.2\%)  & \cellcolor{gray!20}38 (42.7\%) & 82.5\% & 10.3\% \\
    & Method  &  831 (14.3\%) & 32 (36.0\%) & 70.9\% & 14.7\% \\
    & Class   & \cellcolor{gray!20}973 (16.8\%) & 31 (34.8\%) & 67.7\% & 15.5\% \\
\midrule
\multirow{3}{*}{Qwen3-coder}
    & Minimal &  885 (15.3\%) & 32 (36.0\%) & 70.3\% & 14.4\% \\
    & Method  & \cellcolor{gray!20}1028 (17.8\%) & \cellcolor{gray!20}34 (38.2\%) & 68.9\% & 13.4\% \\
    & Class   & \cellcolor{gray!20}1296 (22.4\%) & \cellcolor{gray!20}34 (38.2\%) & 66.5\% & 11.1\% \\
\midrule
\multirow{3}{*}{GPT-OSS}
    & Minimal &   24 (0.4\%)  &  9 (10.1\%) & 98.0\% & 1.6\% \\
    & Method  &  108 (1.9\%)  & 17 (19.1\%) & 96.6\% & 1.6\% \\
    & Class   &  151 (2.6\%)  & 21 (23.6\%) & 96.3\% & 1.1\% \\
\midrule
\multicolumn{2}{@{}l}{\textbf{Average}}
             & 634.9 (11.0\%) & 27.6 (31.0\%) & 79.7\% & 9.3\% \\
\bottomrule
\end{tabular}
\end{table*}

\textbf{Motivation.} 
BreakGuard generates tests for each focal method, i.e., a method in the client code that invokes the target library API. However, LLM-generated tests are not necessarily executable: prior studies have reported compilation errors and execution failures in them \cite{schafer2023empirical, yuan2023chatunitest}. Consequently, a generated test that does not successfully compile and execute on the pre-breaking version cannot be used to detect a breaking change since its failure may be unrelated to the library update. We therefore assess the validity of generated tests before using them for breaking-change detection.

\textbf{Approach.}
For each of the 89 BUMP instances, BreakGuard generates tests for all 5,790 focal methods, for each model-context variant combination.
In this RQ, we report results in terms of valid test coverage at two granularities: (i) \textbf{focal method
coverage} and (ii) \textbf{BUMP instance coverage}. We consider a test to be valid if the test 1) compiles,  2) runs without errors, and 3) passes all assertions on the pre-breaking version. A focal method is covered if BreakGuard generates at least one valid test for it. We compute focal method coverage as the fraction of the 5,790 focal methods for which BreakGuard generates at least one valid test. A BUMP instance is covered if at least one of its focal methods is covered. We compute BUMP instance coverage as the fraction of the 89 instances for which at least one of its focal methods has a valid test.

\textbf{Results.} Table~\ref{tab:rq1_quality} reports the effectiveness of BreakGuard in generating valid tests. We organize the findings around the following takeaways:
\\

\noindent
\textbf{BreakGuard's effectiveness varies significantly per LLM, generating valid tests that cover between 0.4\% and 22.4\% of all focal methods.}
Qwen3-Coder with Class context produces the largest number of valid tests, covering 22.4\% of all focal methods. 
GPT-4o follows at 16.8\% with the same context, while GPT-OSS never exceeds 2.6\%, roughly an order of magnitude below the other two models.
This gap is consistent across all three context variants and holds at both the file and instance levels, indicating that larger models are more successful at generating valid tests.
Even the best configuration therefore leaves more than three-quarters of focal methods without a single usable test.
\\

\noindent
\textbf{Compilation failures are the main cause (avg of 79.7\%) of most invalid tests.}
Compilation failures dominate across all configurations, accounting for 66\% to 98\% of test generation failures. 
Even in the best configuration (Qwen3-coder with the class context), 66.5\% of generated tests do not compile in our experiment. We find that low single-shot compilation rates are consistent with prior work. 
Ouédraogo, Wendkûuni C., et al report high compilation failure rates up to 86\% from a large-scale automated unit test generation study \cite{ouédraogo2026largescaleindependentcomprehensivestudy}. 
Another work by Andrea Lopes et al. also reported low compilation success rates of 36\%, meaning 76\% of generated tests failed to compile \cite{lops2025llmsautomatedunittest}.

\noindent
\textbf{Using GPT-4o and Qwen3 models, BreakGuard generates valid tests for 34\% to 42\% of BUMP instances.}
While it is expected that greater coverage of focal methods would translate into greater coverage of BUMP instances, the results show a more nuanced picture.
GPT-4o with Minimal context produces the fewest valid tests of any GPT-4o configuration (7.2\% of focal methods) yet covers the largest number of BUMP instances (42.7\%). Qwen3-coder with Class context generates more than 3 times more valid tests (22.4\%), but covers a similar number of BUMP instances (38.2\%). 
A configuration can therefore concentrate many valid tests on a few instances, or spread a few valid tests across many, and the instance-level view shows how many BUMP instances BreakGuard achieves more than one valid test.
At last, GPT-OSS trails well behind, covering between 9 and 21 instances (10.1--23.6\%). depending on the chosen configuration. 
\\

\textbf{Class context helps BreakGuard generate valid tests, mainly by reducing compilation errors.} 
Every model covers more focal methods as the context grows richer.
GPT-4o more than doubles its coverage between Minimal and Class context (from 7.2\% to 16.8\%) as its compilation failures fall from 82.5\% to 67.7\%.
Qwen3-coder and GPT-OSS improve monotonically as well, reaching their best coverage with Class context.
When using the Class context, improvements to the Qwen-coder and GPT-OSS focal methods led to better coverage of BUMP instances. 
This is particularly true on GPT-OSS, where the coverage more than doubled, from 9 (10\%) to 21 instances (23\%).
This, however, was not the case for GPT-4o, which showed its lowest level of BUMP instance coverage (34.8\%) when provided with the Class context.

\begin{tcolorbox}[colback=blue!10,colframe=gray!50,title=Answer to RQ1]
BreakGuard's effectiveness depends heavily on the underlying LLM. 
Using the best-performing model and configuration, the approach generates valid tests for 22.4\% of all focal methods in a single shot, covering 34 of the 89 BUMP instances. 
The majority of test generation failures are due to compilation errors, but providing the Class context helps mitigate this issue.   
\end{tcolorbox}

\begin{table*}
\centering
\footnotesize
\caption{Breaking update instances detected per library category, across models and context
variants. \textit{Valid Instance} (RQ1) is the number of BUMP instances covered by at least one
valid test (of 89); the \textit{Detected} column onward (RQ2) reports how many instances of
each category the generated tests detect, with per-category totals given in the header.}
\label{tab:rq2_detection}
\begin{tabular}{@{} ll r r rrrrrrrrr @{}}
\toprule
& & \textbf{RQ1} & \multicolumn{10}{c}{\textbf{RQ2: Detected per library category}} \\
\cmidrule(lr){3-3} \cmidrule(l){4-13}
\textbf{Model} & \textbf{Context}
  & \textbf{\makecell{Valid Inst.\\(N=89)}}
  & \textbf{\makecell{Detected\\(N=89)}}
  & \textbf{\makecell{JSON-\\Libraries}}
  & \textbf{\makecell{Logging\\API}}
  & \textbf{\makecell{App\\frmwk}}
  & \textbf{\makecell{Web\\server}}
  & \textbf{\makecell{Parser}}
  & \textbf{\makecell{Other\\Utilities}}
  & \textbf{\makecell{Mock-\\ing}}
  & \textbf{\makecell{Maven\\Plugins}}
  & \textbf{\makecell{HTTP\\Clients}} \\
& & & & (27) & (26) & (9) & (6) & (6) & (6) & (4) & (3) & (2) \\
\midrule
\multirow{3}{*}{GPT-4o}
    & Minimal & 38 (42.7\%) & 17 (19.1\%) & \cellcolor{myblue!25}5 & \cellcolor{myblue!45}9 & -- & -- & \cellcolor{myblue!5}1 & \cellcolor{myblue!5}1 & -- & -- & \cellcolor{myblue!5}1 \\
    & Method  & 32 (36.0\%) & 13 (14.6\%) & \cellcolor{myblue!20}4 & \cellcolor{myblue!40}8 & -- & -- & -- & -- & -- & \cellcolor{myblue!5}1 & -- \\
    & Class   & \cellcolor{rowgrey}31 (34.8\%) & \cellcolor{rowgrey}27 (30.3\%) & \cellcolor{myblue!40}8 & \cellcolor{myblue!80}16 & -- & -- & \cellcolor{myblue!5}1 & -- & -- & \cellcolor{myblue!5}1 & \cellcolor{myblue!5}1 \\
\midrule
\multirow{3}{*}{Qwen3-coder}
    & Minimal & 32 (36.0\%) & 18 (20.2\%) & \cellcolor{myblue!15}3 & \cellcolor{myblue!55}11 & \cellcolor{myblue!5}1 & -- & \cellcolor{myblue!5}1 & -- & -- & \cellcolor{myblue!5}1 & \cellcolor{myblue!5}1 \\
    & Method  & 34 (38.2\%) & 20 (22.5\%) & \cellcolor{myblue!25}5 & \cellcolor{myblue!65}13 & -- & -- & \cellcolor{myblue!5}1 & -- & -- & -- & \cellcolor{myblue!5}1 \\
    & Class   & 34 (38.2\%) & 20 (22.5\%) & \cellcolor{myblue!35}7 & \cellcolor{myblue!55}11 & -- & -- & \cellcolor{myblue!5}1 & -- & -- & -- & \cellcolor{myblue!5}1 \\
\midrule
\multirow{3}{*}{GPT-OSS}
    & Minimal &  9 (10.1\%) &  4  (4.5\%) & -- & \cellcolor{myblue!15}3 & -- & -- & \cellcolor{myblue!5}1 & -- & -- & -- & -- \\
    & Method  & 17 (19.1\%) &  8  (9.0\%) & -- & \cellcolor{myblue!40}8 & -- & -- & -- & -- & -- & -- & -- \\
    & Class   & 21 (23.6\%) & 10 (11.2\%) & \cellcolor{myblue!5}1 & \cellcolor{myblue!40}8 & -- & -- & -- & -- & -- & -- & \cellcolor{myblue!5}1 \\
\bottomrule
\end{tabular}
\end{table*}

\subsection{RQ2: \rqtwo}
\label{sec:rq2_results}
\textbf{Motivation.} Automated detection of breaking changes is challenging because library updates can alter API signatures as well as the behavior of existing APIs \cite{client-lib-compatibility-testing, UnderstandingImpact}. Detecting such changes therefore requires exercising library functionalities that are actually used by the client and checking whether its expected behavior is preserved across versions \cite{ochoa2022breaking, RegressionTesting}
While RQ1 established the validity of BreakGuard tests on the pre-breaking version, detecting a breaking change requires a valid test to exercise the affected library functionality and fail on the breaking version.
Therefore, we evaluate BreakGuard's effectiveness as an automated tool for detecting breaking changes, allowing developers to identify the risks of library migration ~\cite{Venturini_2023, cox2019surviving}. 
Furthermore, we analyze the library categories for which BreakGuard is more effective to understand when generated tests provide value and where the approach remains limited.

\textbf{Approach.} We run all valid tests for each breaking update instance on the breaking version. A breaking change is considered \emph{detected} if at least one focal method from a BUMP instance has a valid test on the pre-breaking version but fails on the breaking version. 
To measure how varied these detections are, we also present the results by library categories (as shown in Table \ref{tab:library-scope}) that the successfully detected instances cover.

 \begin{figure}
    \centering
    \includegraphics[width=.5\textwidth]{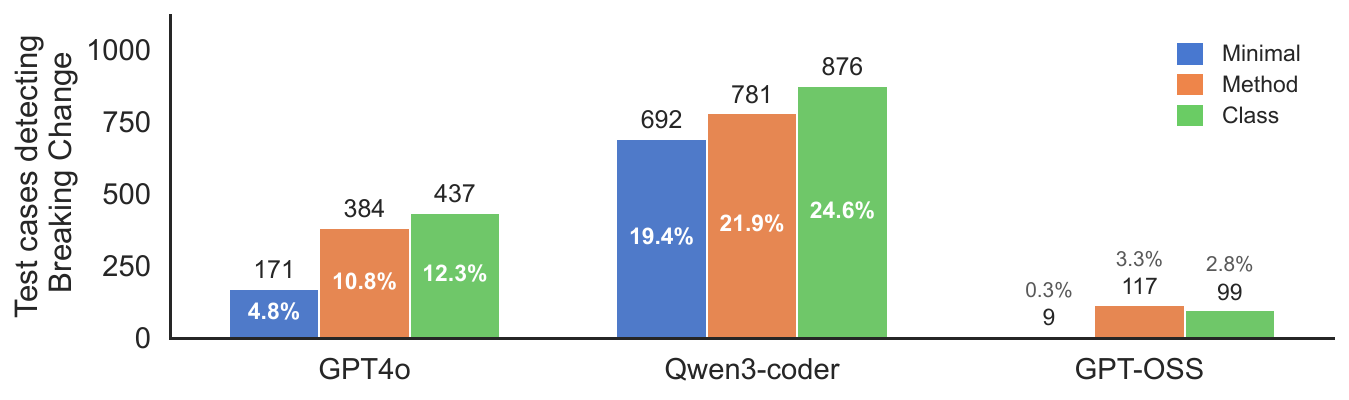}
    \caption{Breaking changes detected when valid tests from Table~\ref{tab:rq1_quality} run against the breaking library version. (Note: each test covers different focal methods)}
    \label{fig:breaking-testcount}
\end{figure}

\begin{figure}
    \centering
    \includegraphics[width=.5\textwidth]{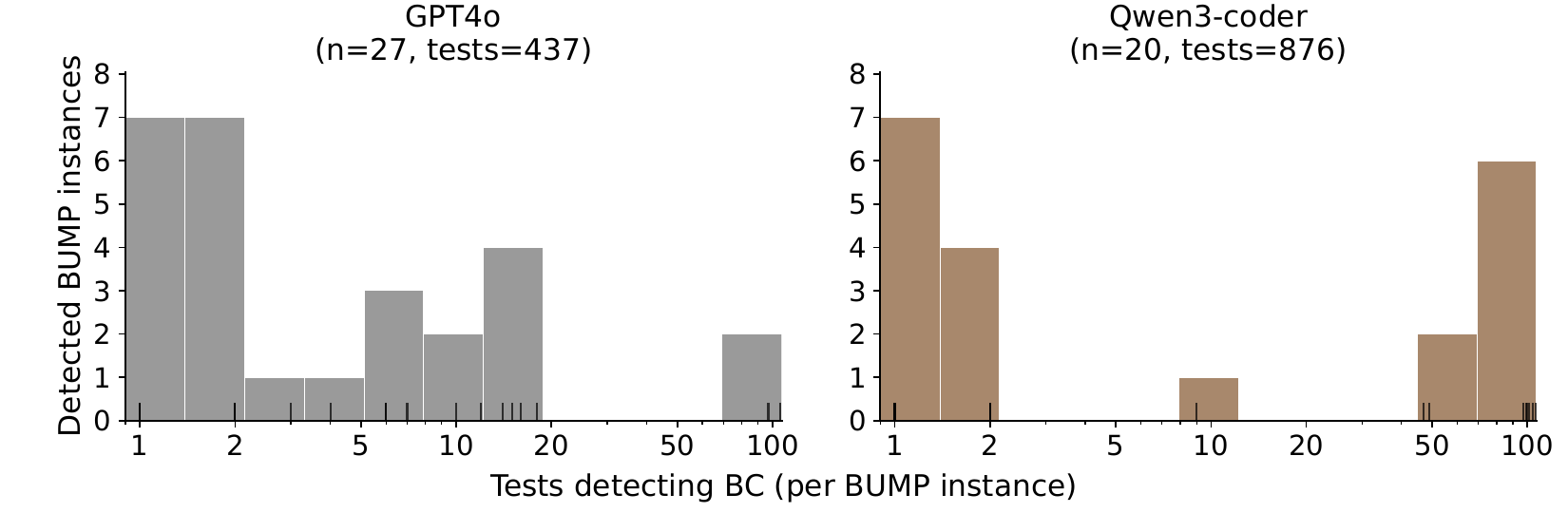}
    \vspace{-0.5em}
    \caption{Distribution of focal-method counts per BUMP instance under the Class context, for GPT-4o and Qwen3-coder, highlighting the reasons for less BC detection even with more focal-method coverage.}
    \label{fig:breaking-distribution}
    \vspace{-1.5em}
\end{figure}

\textbf{Results.} Table~\ref{tab:rq2_detection} reports the outcomes across 89 BUMP instances. We organize the findings around the following takeaways.
\noindent
\textbf{Tests generated by BreakGuard detect breaking changes in up to 30.3\% of the instances, in a single shot.}
Tests generated by GPT-4o with the Class context perform best, detecting the break in 27 of the \Ninst instances (30.3\%).
Using the Class context, Qwen3-coder's generated tests detect up to 20 (22.5\%) instances, while GPT-OSS's detects at most 10 breaking changes (11.2\%).
BreakGuard detects nearly one in three breaking updates while generating each test in a single shot, from the client code alone, without any knowledge of the update that will break it.

\textbf{GPT-4o tests were more successful at detecting breaking changes, despite Qwen3-coder's initially better coverage.}
Tests generated by GPT-4o with the Class context detect 27 out of the 31 covered BUMP instances. 
On the same context configuration, Qwen3-coder cover more instances (34) but tests were effective in only 20 BUMP instances. 
As seen in Figure~\ref{fig:breaking-testcount}, due to its high volume of generated tests, Qwen3-coder tests also fail more often. In total, 873 of Qwen3-coder's tests fail on the breaking version of BUMP instances, compared to 437 tests by GPT-4o. 
To understand the number of failing tests per BUMP instance, we report their distribution in Figure \ref{fig:breaking-distribution}.
As shown, Qwen3-coder generated more than 50 failing tests in the breaking version across 7 distinct BUMP instances. 
GPT4-o, on the other hand, had most BUMP instances detected with fewer than 20 failing tests. 
There is certainly a potential trade-off to consider in the practical application of BreakGuard: Qwen3-coder's multiple failing tests can increase confidence in a reported breaking change; however, the GPT-4o test suite was more effective at detecting the break across more BUMP instances.

\begin{figure}
    \centering
    \includegraphics[width=.5\textwidth]{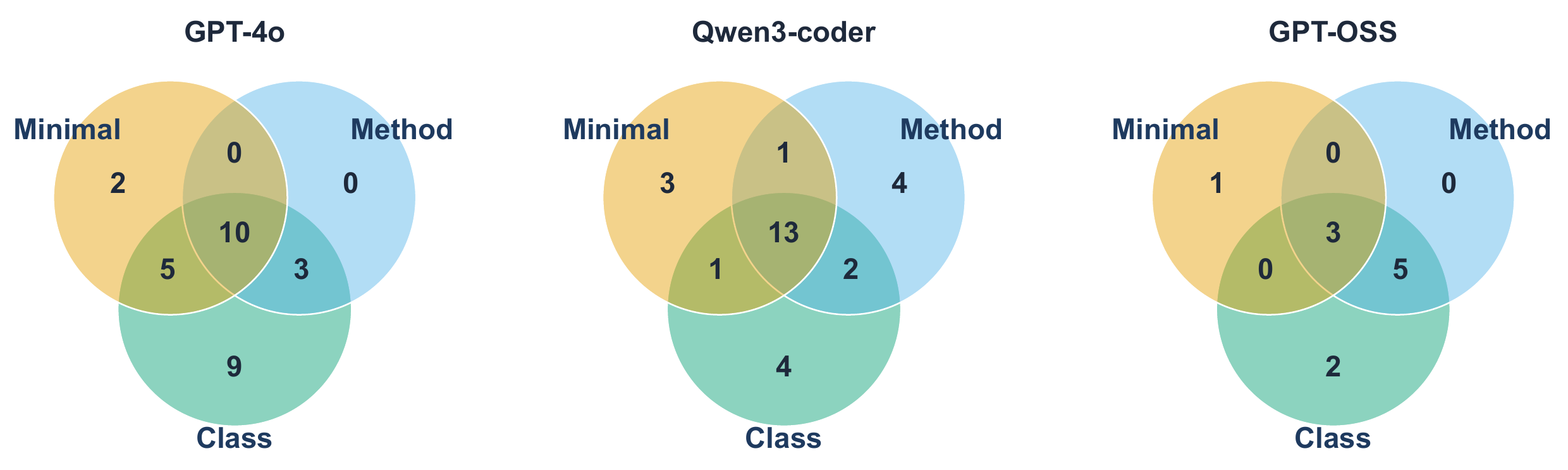}
    \vspace{-0.5em}
    \caption{Breaking change detection overlap across different context variants and models.}
    \vspace{-1.5em}
    \label{fig:venn_context}
\end{figure}

\textbf{Tests generated with the Class context detect more breaking changes, yet every context variant detects unique breaking updates.}
Tests generated by GPT-4o improve from 17 detected instances under Minimal context to 27 under Class context. Qwen3-Coder detects between 18 and 20 across all three. Figure~\ref{fig:venn_context} shows BUMP instance detection overlap across different contexts through a Venn diagram. 
Most breaking changes are detected by more than one variant of BreakGuard context. Overall, the Class context finds more unique breaking changes; however, other variants also contribute to unique detections depending on the model. Except for GPT4o and GPT-OSS, we found that method variants do not contribute to any unique detections.   

\textbf{BreakGuard detects breaking changes across multiple library categories.}
Table \ref{tab:rq2_detection} presents the detection count across different client and library categories. 
We find that, while GPT-4o with Class context achieves the highest detection count (27 instances), its detections cover 5 different library categories, namely (JSON libraries, Logging, Parser generator, Maven Plugins, and HTTP Clients). 
In contrast, Qwen3-coder under Minimal context achieves just 18 detections but covers 6 different library categories. 
However, detections per category are uneven: Logging API is the only category identified by all nine configurations, while Mocking is detected only by GPT4o and Minimal. 
Detections concentrate in the two categories, namely Logging API and JSON Libraries, that are largely represented in the dataset \ref{tab:bump_categories}. 
The remaining categories are detected only sparsely. For example, Application framework and Data structures appear under only a single Qwen3-coder configuration. One reason we found is that the most frequently detected library categories also have more valid focal methods; the LLM generated more tests to capture breaking changes in them.

\begin{tcolorbox}[colback=blue!10,colframe=gray!50,title=Answer to RQ2]
BreakGuard successfully detects up to 27 out of 89 (30.3\%) breaking changes, with tests generated by GPT-4o and up to 20 out of 89 (22\%) with tests generated by Qwen3-coder. 
Providing the Class context to LLMs improves effectiveness in detecting breaking changes. Detections cover different library categories but are mostly concentrated in the two most represented categories in the dataset, namely Logging and JSON Libraries.
\end{tcolorbox}

\subsection{RQ3: \rqthree}
\label{sec:rq3_results}
\textbf{Motivation.}
In this RQ, we aim to understand what failure types BreakGuard detects and how generated tests reveal breaking changes. 
A detected breaking change may appear as a runtime error caused by a structural change, such as a missing class or method, or as an assertion failure that captures a behavioral change \cite{client-lib-compatibility-testing}. 
This characterization helps practitioners understand what BreakGuard
tests reveal after a dependency update and helps researchers identify improvements in LLM-generated tests for breaking change detection.

\textbf{Approach.}
For every generated test that identifies a breaking change, we analyze the test failure in two steps.
First, we obtain the execution outcome from Maven Surefire.%
It reports whether each test fails due to an assertion violation (\texttt{FAILURE}) or terminates due to an unexpected exception (\texttt{ERROR}). A \texttt{FAILURE} means that an expected behavior checked by the test did not hold, whereas an \texttt{ERROR} means that an exception prevented the test from completing.

Second, we wrote a script to characterize each breaking failure type. For every \texttt{ERROR}, the script records the root-cause exception by parsing the text around (the \texttt{Caused by:}) to characterize the runtime failures. This approach is commonly used in failure analysis because the root-cause exception provides the most specific information about the underlying failure rather than the intermediate exceptions propagated through the call stack. \cite{UnderstandingImpact, BehaviouralBackwardCompatibility}.
Similarly, for every \texttt{FAILURE}, we inspect the assertions generated  by the LLM. We distinguish assertions that check program values from those that only verify successful execution, such as \texttt{assertDoesNotThrow}, and other assertion patterns. We report all results at the test-case level.

\begin{table}
\centering
\caption{Test failure types across model and context variant, split into ERROR and FAILURE. \textbf{Tests} counts individual test cases that detect a breaking change; \textbf{BUMP Instance} counts the distinct BUMP instances in which at least one such test captures the breaking change.}
\label{tab:rq3-mech}
\begin{tabular}{ll rr rr}
\toprule
& & \multicolumn{2}{c}{ERROR} & \multicolumn{2}{c}{FAILURE} \\
\cmidrule(lr){3-4}\cmidrule(lr){5-6}
\textbf{Model} & \textbf{Context} & \textbf{Tests} & \textbf{\makecell{BUMP\\Instance}} & \textbf{Tests} & \textbf{\makecell{BUMP\\Instance}} \\
\midrule
\multirow{3}{*}{GPT-4o}
 & Minimal & 169 & 15 &  2 &  2 \\
 & Method  & 384 & 13 &  0 &  0 \\
 & Class   & 422 & 18 & 15 &  9 \\
\midrule
\multirow{3}{*}{Qwen3-coder}
 & Minimal & 611 & 11 & 81 &  7 \\
 & Method  & 729 & 13 & 52 &  7 \\
 & Class   & 867 & 14 &  9 &  6 \\
\midrule
\multirow{3}{*}{GPT-OSS}
 & Minimal &   9 &  4 &  0 &  0 \\
 & Method  & 107 &  8 & 10 &  3 \\
 & Class   &  92 &  9 &  7 &  1 \\
\midrule
\multicolumn{2}{l}{Total} & 3{,}390 & -- & 176 & -- \\
\bottomrule
\end{tabular}
\end{table}

\begin{table}
\centering
\caption{How valid LLM-generated tests fail on the breaking version
(test level), reporting the root cause}
\label{tab:rq3-mech-type}
\begin{tabular}{lrr}
\toprule
Outcome & Test cases & \% \\
\midrule
\textbf{(\texttt{FAILURE})} & 176 & 4.9 \\
\quad AssertionFailedError  & 163 & 4.6 \\
\quad AssertionError        & 5   & 0.1 \\
\quad Other         & 8   & 0.2 \\
\midrule
\textbf{ (\texttt{ERROR})}       & 3{,}390 & 95.1 \\
\quad NoClassDefFoundError   & 1{,}441 & 42.5 \\
\quad NoSuchMethodError     & 984 & 29.0 \\
\quad ClassNotFoundException & 965 & 28.5 \\
\midrule
Total & 3{,}566 & --\\
\bottomrule
\end{tabular}
\end{table}

\begin{table}
\centering
\caption{The 176 assertion detections, grouped by the kind of assertion the
model wrote.}
\label{tab:rq3-oracle}
\begin{tabular}{lrr}
\toprule
Assertion kind & Test cases & \% \\
\midrule
Should not throw (e.g. \texttt{assertDoesNotThrow}) & 136 & 77.3 \\
Value oracle (e.g.\ \texttt{assertTrue} on output)   & 25 & 14.2 \\
Existence probe (\texttt{assertNotNull}/\texttt{fail}) & 7 & 4.0 \\
No explicit assertion                                & 8 & 4.5 \\
\midrule
Total & 176 & -- \\
\bottomrule
\end{tabular}
\end{table}

\textbf{Results.} Tables~\ref{tab:rq3-mech}, \ref{tab:rq3-mech-type}, and \ref{tab:rq3-oracle} summarize the following three analyses: 
1) the types of failures reported by the tests, 2) the root cause of tests errors and 3) the assertion strategies in the assertion failures.

\textbf{BreakGuard identifies breaking changes primarily through runtime errors.}
Table~\ref{tab:rq3-mech} shows the execution outcomes of tests that identify breaking changes, across all models and configurations.
Among the 3{,}566 detecting tests, 3{,}390 (95.1\%) terminate with a Maven Surefire \texttt{ERROR}, while only 176 (4.9\%) terminate with a \texttt{FAILURE}. 
This pattern holds across models and context variants. For example, GPT-4o produces only 17 assertion failures, whereas Qwen3-coder produces 142 of the 176 assertion failures. 

\textbf{Most runtime errors are caused by missing classes or methods.}
For the 3{,}390 tests reported as \texttt{ERROR}, Table~\ref{tab:rq3-mech-type}
reports the root-cause exceptions extracted from their stack traces. All
observed runtime errors are caused by missing classes or methods after the
dependency update. The dominant causes are \textbf{NoClassDefFoundError}
(1{,}441; 42.5\%), \textbf{NoSuchMethodError} (984; 29.0\%), and
\textbf{ClassNotFoundException} (965; 28.5\%). This differs from developer-written tests in BUMP, where runtime errors also include exceptions such as IllegalStateException and ClassCastException ~\cite{reyes2024bump}. These results indicate that BreakGuard primarily exposes structural breaks.

\textbf{Failure-based detections rarely check behavioral changes.}
Table~\ref{tab:rq3-oracle} analyzes the 176 tests reported as
\texttt{FAILURE} based on the assertion strategies generated by the models.
Most of these cases (136; 77.3\%) use assertions such as
\texttt{assertDoesNotThrow}, which verify that a method call completes successfully. 
Although Surefire reports these tests as assertion failures, the underlying cause is often still a runtime error caused by the dependency update. 8 tests (4.5\%) in the other category captured as FAILURE use the TestNG framework, which puts runtime errors and assertion failures in the same bucket. We reviewed each one manually and found that they are runtime errors. 
Only 25 tests (14.2\% of assertion failures and 0.7\% of all detecting tests) use value-based assertions that check behavior. This result is expected in our study. Although we ask the LLM to cover nominal, minimal, boundary, and error scenarios \ref{fig:promptdesign} and provide the focal methods, their API call sites, and the surrounding code, we do not explicitly provide expected output and behavioral specifications. Thus, the LLM has to infer the expected behavior from the provided source code and API usages, making execution-based assertions more likely than value-based assertions.

\begin{tcolorbox}[colback=blue!10,colframe=gray!50,title=Answer to RQ3]
Using only static information from clients, BreakGuard effectively detects crash-type breaking changes through runtime errors, accounting for 95.1\% of detection cases across all models. It also captures behavioral changes through value-based assertions in 25 out of 176 tests. 
\end{tcolorbox}

\subsection{RQ4: \rqfour}

\textbf{Motivation.} LLM-based test generation introduces inference costs that can become substantial when tests are generated at scale. Prior large-scale studies have generated hundreds of thousands of tests using LLMs \cite{ouédraogo2026largescaleindependentcomprehensivestudy, chu2026llmtestsurvey}. Since BreakGuard generates tests for each client focal method that calls library methods, its cost depends on the number of API usages and the amount of context provided to the model. We therefore analyze how these factors affect the token usage to help practitioners estimate the cost of applying BreakGuard and choose an appropriate context configuration.

\textbf{Approach.} We record the input and output tokens for every generated test file under each (model, context) pair. We count tokens using each model's native tokenizer: tiktoken with the \texttt{o200k\_base} encoding for GPT-4o and Hugging Face
AutoTokenizer for Qwen3-480B and GPT-OSS-120B. We aggregate token counts at both the focal method and the breaking-update-instance level. Finally, we estimate the mean and median dollar cost per BUMP instance using the OpenRouter price list available on July 26, 2026~\cite{openrouter}.

\begin{table}
\centering
\scriptsize
\setlength{\tabcolsep}{3.5pt}
\renewcommand{\arraystretch}{1.2}
\caption{Token cost and estimated dollar cost of BreakGuard across models and context levels. Per Focal Method (FM) values are mean tokens; per BUMP instance values are mean and median token counts.}
\label{tab:cost}
\begin{tabular}{@{}ll rrr rr rr@{}}
\toprule
& & \multicolumn{3}{c}{\textbf{FM (mean)}}
  & \multicolumn{2}{c}{\textbf{BUMP Instance}}
  & \multicolumn{2}{c}{\textbf{USD/detection.}} \\
\cmidrule(lr){3-5} \cmidrule(lr){6-7} \cmidrule(l){8-9}
\textbf{Model} & \textbf{Context}
  & \textbf{In} & \textbf{Out} & \textbf{Total}
  & \textbf{Mean} & \textbf{Median}
  & \textbf{Mean} & \textbf{Median} \\
\midrule
\multirow{3}{*}{GPT-4o}
  & Minimal & 894   & 225 & 1{,}119 & 72{,}785  & 11{,}942 & 0.19 & 0.046 \\
  & Method  & 1{,}248 & 378 & 1{,}626 & 105{,}750 & 15{,}352 & 0.28 & 0.065 \\
  & Class   & 3{,}908 & 413 & 4{,}321 & 281{,}117 & 27{,}583 & 0.90 & 0.088 \\
\addlinespace
\multirow{3}{*}{Qwen3-480B}
  & Minimal & 909   & 378 & 1{,}287 & 83{,}750  & 12{,}937 & 0.026 & 0.008 \\
  & Method  & 1{,}264 & 572 & 1{,}836 & 119{,}415 & 18{,}113 & 0.040 & 0.013 \\
  & Class   & 3{,}886 & 489 & 4{,}376 & 284{,}662 & 30{,}641 & 0.065 & 0.014 \\
\addlinespace
\multirow{3}{*}{GPT-OSS-120B}
  & Minimal & 894   & 384 & 1{,}278 & 83{,}145  & 11{,}883 & 0.005 & 0.001 \\
  & Method  & 1{,}248 & 604 & 1{,}851 & 120{,}450 & 17{,}068 & 0.008 & 0.001 \\
  & Class   & 3{,}908 & 640 & 4{,}548 & 295{,}852 & 29{,}732 & 0.011 & 0.002 \\
\bottomrule
\end{tabular}
\end{table}

\textbf{Results.} Table \ref{tab:cost} presents the token cost per focal method and per instance across all model and context variant pairs. 

\textbf{Class context requires roughly 3--4 times more tokens than minimal context.} 
Moving from minimal to class context increases the total tokens per focal method from 1{,}119 to 4{,}321 tokens for GPT4o, from 1{,}287 to 4{,}376 for Qwen3-480B, and from 1{,}278 to 4{,}548 for GPT-OSS-120B. This increase is driven primarily by input tokens, which are largely determined by the context provided and remain similar across models. Output tokens vary by model but remain a small share of the total under richer contexts.

\textbf{BreakGuard costs less than \$0.09 per median BUMP instance across configurations, but costs increase substantially for clients using more API call-sites.} 
Across model and context configurations, the median cost per BUMP instance ranges from \$0.001 to \$0.088, while the mean ranges from \$0.005 to \$0.90. The difference between the mean and median reflects the concentration of token consumption among instances, with many focal methods having API call sites (Figure \ref{fig:APICALLSIGHTSTATS}. We see that the median instance requires test generation for 12 focal methods; however, 15 instances contain more than 100 call sites, and 3 contain more than 500 in our dataset. Thus, the cost of applying BreakGuard is generally low for a typical instance but can increase substantially for projects with many library API usages.

\textbf{Higher-cost configurations provide higher detection in some cases, but the absolute cost remains small.}
Combining cost with detection results from RQ2, the highest-cost setup is only sometimes the best. GPT-4o reaches the top detection rate (30.3\%, 27/89) only with class context, at a median cost of roughly \$0.088/instance. At minimal and method context, it detects fewer breaks
than Qwen3-coder-480B. Whereas Qwen3-coder-480B detects 22.5\% (20/89) at method context and gains nothing from the richer class context, its most effective configuration costs roughly \$0.013/BUMP instance, which is roughly one-seventh of GPT-4o's cost but at a loss of seven breaking-change detections. Nevertheless, even the highest-performing configuration costs only \$0.088 per median BUMP instance, suggesting that monetary cost is unlikely to be a substantial barrier to applying BreakGuard.
\noindent
\begin{tcolorbox}[colback=blue!10,colframe=gray!50,title=Answer to RQ4]

The mean cost of BreakGuard ranges from \$0.005 to \$0.90 USD per breaking change across model and context variants. Cost is primarily determined by the number of focal methods used for test generation and the context provided to the LLM. Richer context increases token consumption by roughly 3--4 times per focal method; however, the best configuration costs only \$0.90 per breaking change detection, making BreakGuard a feasible option for breaking Change detection.
 
\end{tcolorbox}

\section{Discussions and Implications}
\label{sec:discussion}

Our findings reveal both the potential and limitations of using LLMs to generate tests for breaking-change detection. In this section, we discuss the key insights, their implications for researchers and practitioners, and the limitations of our study design.

\subsection{Why generated tests fail to detect breaking changes?}

RQ2 focuses on reporting the effectiveness of generated tests in detecting breaking changes. 
However, a significant number of all valid tests also pass on the breaking version. 
To understand why the tests miss detecting a breaking change, we particularly investigate tests from BUMP instances where detection failed entirely (meaning none of the valid tests could detect breaking change). This covers 2{,}111 test cases across 89 instances, 3 models, and
3 context variants. To inspect this scenario, we applied a simple logic: when a test exercises the broken call API method, its class must load before its methods can run; if the affected API class never loads while a test executes, then that test never exercised the break. Since the default Maven test scope does not provide a detailed execution trace for a passing test. we rerun all successful tests on the breaking version with \texttt{-verbose:class} instrumentation to check whether the broken API class ever loaded. 
We found 1{,}656 (78.4\%) tests never loaded the broken API class at all, so these tests could not have exercised the break. The remaining 455 (21.6\%) tests loaded the broken API class but passed on the breaking version. We manually inspected 80 of these 455 tests ( sampled with a $\pm$10\% margin of error, 95\% confidence).  For every 80 tests, we inspected the code of the generated test, its execution log, and the prompt context passed to the LLM to understand the root cause, and we found three patterns.

\textbf{A silent null-object fallback hides the break.} 
In 37 of 80 tests, the model called the affected method with correct inputs, but the test still passed on the breaking version as the break resides in a transitive dependency whose context is not passed to the model. 
One example is the SLF4J library upgrading from 1.7.36 to 2.0.2. This upgrade replaces the old static-binding lookup with a new provider mechanism in the library code that doesn't recognize the client's existing logging binding, so SLF4J silently falls back to a null-object implementation (\texttt{NOPLogger}) that discards every call rather than throwing an error. The test reached the broken API method \texttt{logger.info()}, yet it passed because the prompt never describes any runtime information to assert on or to guide the change in the library. 

\textbf{Generated test coverage is insufficient to detect the break} 
In 32 of 80 tests, the model constructs the right object but calls the broken API method inside a catch block. This causes the test to silence the thrown exception. 
In another case, the test calls another API method from the affected API class. The class loads, but the test does not reach the breaking path.

\begin{figure}
    \centering
    \includegraphics[width=.5\textwidth]{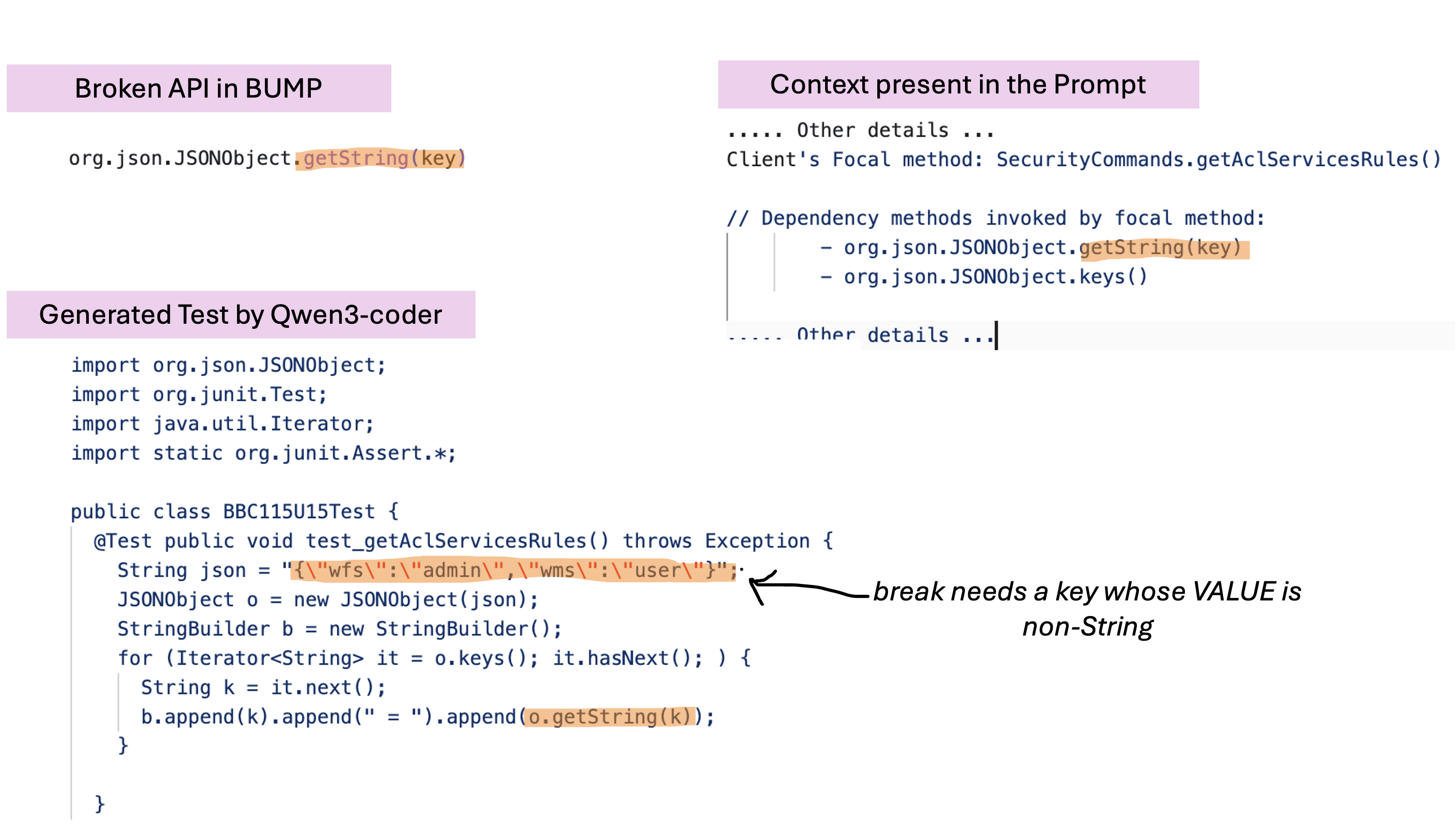}
    \caption{A weak input due to which a breaking change was undetected:
    \texttt{getString()} is called only on String values, while the break
    triggers on an Integer.}
    \label{fig:wrongInput}
\end{figure}

\textbf{The inputs used by LLMs are insufficient to detect the break} 
In the remaining 11 of 80 tests, the model calls the affected method but cbut only with inputs that do not exercise the change. Figure~\ref{fig:wrongInput} shows one: the model called \texttt{getString()} on the \texttt{org.json} library, but only with String-typed values. The break triggers when \texttt{getString()} runs on an Integer value.

Across all three patterns, the test compiles, executes, and passes in both the pre-breaking and breaking versions. This is a fundamentally different problem that provides no failure signal; it is an ineffective test, giving nothing to correct against. Therefore, it cannot benefit by simply applying agentic pipeline or iterative loops. Rather, fixing it requires dependency-related information the LLM didn't have in the first place. For example, what the change actually does,  exactly which call site it affects, and what runtime conditions the client uses while exercising these call sites.

\subsection{Feedback loops can improve compilation failures}
\label{subsec:agentic}

From RQ1, we find that even with the best-performing configuration (Qwen3-coder with class context), BreakGuard only generated 33.5\% of tests that compile. To understand the high rate of compilation failures,  we looked into the execution log of 3,851 out of 5790 failing test files from Qwen3-coder with class context. 
We developed a simple parser following the compilation error taxonomy of Mesbah et al.~\cite{10.1145/3338906.3340455} to categorize the error types.
From this analysis, we find two error types dominate: undefined symbols (cant.resolve) and undefined packages (doesn't.exist), which account for 62\% and 22\% of failures, respectively. These results are consistent with prior studies showing that LLM-generated tests frequently fail to compile because of hallucinated API references, missing imports, and incorrect dependencies \cite{ouédraogo2026largescaleindependentcomprehensivestudy, chu2026llmtestsurvey, berndt2026flakiness}. 

These failures are potentially recoverable because the compiler provides concrete information about what went wrong. Prior work has shown that feeding compilation errors back to the LLM can enable it to repair failing tests~\cite{xu2025hallucination}. Our study deliberately uses single-shot prompting without such feedback loops to provide a baseline on what an LLM could achieve in test generation that can capture breaking changes. Thus, the observed low compilation rate represents a lower bound on what could be achieved with iterative loop or agentic repair.

\subsection{Generated tests frequently write assertions but often implement exception handling that swallows the break }

We also observe that LLM-generated tests include \texttt{assert} statements approximately 80\% of the time. However, LLMs also tend to wrap focal method or API calls inside \texttt{try-catch} blocks. While this mirrors general software engineering practices to prevent system crashes, it actively works against the goal of testing. When an API undergoes a breaking change that manifests as an exception or runtime failures, the try-catch block will either block the error before it can fail, bypassing the assertions, or land inside the catch, which does not check anything significant. This pattern directly suppresses detection and becomes a silently passing test rather than a detection signal. Future approaches could explicitly instruct LLMs to let unexpected exceptions propagate as test failures rather than catching them.

\subsection{Implications for Practitioners}

Based on our findings, we offer the following observations for practitioners considering LLM-based approaches for breaking change detection:

\textbf{BreakGuard is good at detecting test cases that detect crash-type breaking changes.}
In its current form, BreakGuard is most effective at detecting breaking changes that manifest as missing classes or methods. Although behavioral breaking-change detection is less effective, practitioners can still reuse the generated test suites from BreakGuard to proactively test future library upgrades.

\textbf{Providing Class-level context improves the effectiveness of generated tests.} 
Among the configurations we evaluate, full focal-class context provides the best trade-off between detection effectiveness and cost. We find that providing richer context gives the LLM surrounding logic to generate tests with better assertions. As discussed above, practitioners could further enrich context with runtime and other dependency-related information to detect more breaking changes. 

\textbf{Generated tests should avoid suppressing exceptions that may reveal breaking changes.} 
We find that LLMs frequently wrap library calls in \texttt{try-catch} blocks, suppressing exceptions which could otherwise signal a breaking change. Practitioners should therefore constrain prompts or apply lightweight post-processing to remove overly broad exception handling, which could improve detection rates at low cost.

\subsection{Implications for Researchers}

Our study opens several research directions:

\textbf{Evaluate feedback-driven test generation for breaking update detection} Our single-shot results provide a baseline. Future work should evaluate iterative repair and multi-turn generation approaches on the same BUMP instances to quantify the improvement.

\textbf{Develop new strategies for breaking change detection with LLMs.} Our cost analysis results in RQ4 show that Cost and effort concentrate in a few clients with hundreds of call sites (Fig.~\ref{fig:APICALLSIGHTSTATS}), making reducing the number of generated tests an attractive efficiency target. Simply deduplicating the same call sites is risky, as they may be invoked with different arguments, sit in different control-flow contexts, and therefore exercise different behavior. Additionally, a breaking change might surface at one call site but not another. Future approaches could use library code changes and dependency-aware call graphs to prioritize call sites whose reachable behavior is affected, while also accounting for dynamic information in the client code.

\textbf{Cross-ecosystem generalization.} Our study focuses exclusively on Java/Maven projects. Whether these findings generalize to other ecosystems (npm/JavaScript, pip/Python, Cargo/Rust) with different dependency resolution mechanisms and testing conventions remains an open question.

\section{Related Works}
\label{sec:related}

We situate our work at the intersection of three research areas: breaking change detection and test adequacy, Regression testing for dependency compatibility, and LLM-based test generation. To the best of our knowledge, no prior study has evaluated the effectiveness of LLM-generated tests in detecting breaking changes caused by library version updates. We summarize representative work in each area and highlight the gap our study addresses.

\subsection{Breaking Change Detection and Test Adequacy}
\label{sec:rw_bc_detection}

Approaches to detecting breaking changes in library updates generally fall into two categories: \textbf{static} and \textbf{dynamic}. Static approaches compare API surfaces across library versions without executing code. Tools such as Clirr~\cite{clirr}, JApiCmp~\cite{simons2023japicmp}, and Maracas~\cite{ochoa2022breaking} detect removed methods and signature changes. However, static approaches are limited to syntactic differences and cannot reliably detect behavioral breaking changes where API signatures remain unchanged~\cite{gyori2016crosschecker, jiang2018detecting}. Dynamic approaches execute client tests against both the old and new library versions, identifying breaking changes when tests pass on the old version but fail on the updated dependency. The BUMP benchmark~\cite{reyes2024bump}, which we use as ground truth, follows this paradigm. Dynamic detection captures both syntactic and behavioral regressions but is fundamentally constrained by the quality and coverage of client test suites. Hejderup and Gousios~\cite{hejderup2022trust} show that many dependency APIs used by client projects are never exercised by their tests. On the other hand, Raj and Costa~\cite{raj2025supportingopensourcelibrary} found that OSS library projects also lack testing for many API calls that their clients depend on, leaving breaking changes undetected. At the same time, studies of library evolution report that breaking changes remain common despite semantic versioning practices~\cite{xavier2017historical, raemaekers2017semantic, jezek2015apis}. 

\textbf{Gap.} Existing static techniques detect signature-level changes and produce reports for the whole library that are often noisy from the perspective of an individual client, as a client only uses a few of their API methods \cite{harrand2022api}. Dynamic techniques detect behavioral regressions only where suitable tests already exist. Our work bridges these limitations by statically identifying client-side library API call sites and using them to generate targeted tests that explicitly cover the dependency APIs. These enhanced test suites can also serve as a migration test suite for clients in future updates.

\subsection{Regression Testing for Dependency Compatibility}
\label{sec:rw_test_adequacy}
Research shows that updating library versions in the client code often introduces test failures \cite{ochoa2022breaking}. Xavier et al. show that approximately 15\% of library updates introduce breaking changes ~\cite{xavier2017historical}. To mitigate the risks of breaking changes, client write tests to validate their own logic, which indirectly exercises the logic of the OSS library they use. These tests are regression tests which ensure that changes do not alter their software's behavior \cite{RegressionTesting}. However, Jayasuriya et al. note that even when client tests trigger failures, the distance between the root cause and the failing assertion severely hinders diagnosis and remediation \cite{UnderstandingImpact}.  To address this issue, developers sometimes write tests that directly exercise library APIs and focus on verifying outputs and side effects \cite{client-lib-compatibility-testing}. However, client tests primarily concern validating their own logic, and writing such library-focused tests is relatively rare in practice.

\textbf{Gap.} Prior work establishes that test coverage of library APIs is insufficient and that breaking changes are common, but no existing approach automatically generates tests to close this gap. Our study directly addresses this by generating targeted tests from every OSS API usage on the client side.

\subsection{LLM-based Test Generation}
\label{sec:rw_llm_test}

Recent work has shown that LLMs can generate unit tests with meaningful coverage \cite{lops2025llmsautomatedunittest, yuan2023chatunitest}. Beyond initial generation, recent work also explores integrating static analysis to improve generated tests and assertions~\cite{xu2025hallucination, roychowdhury2025staticprogramanalysisguided}. These papers compare LLM-generated tests with traditional approaches such as EvoSuite~\cite{fraser2011evosuite} and Randoop~\cite{pacheco2007randoop} to assess structural coverage and assertion generation. Additionally, Nan et al. demonstrate that enriching prompts with statically extracted program information substantially improves generated tests ~\cite{Nan2025Test}. We adopt their approach to construct structured prompts from statically extracted client-side information, including focal methods, library API call sites, argument expressions, and type references. Recent work also applies LLMs to repair already detected breaking changes. Byam et al. use structured prompts containing compiler errors and API differences to fix compilation failures in client code due to version updates ~\cite{reyes2026byamfixingbreakingdependency}. Fruntke and Krinke investigate iterative prompting and agentic repair workflows on the BUMP benchmark ~\cite{fruntke2025fixing}. Both studies demonstrate that richer contextual information improves repair effectiveness.

\textbf{Gap.} Existing LLM-based test generation work primarily aims to improve code coverage or detect faults in the program under test. In contrast, LLM-based dependency-update work has focused on repairing incompatibilities after a breaking update has already been identified. Neither investigates whether LLM-generated tests from client–library usage can proactively detect breaking changes in clients while updating to a newer library version. To our knowledge, our work is the first to evaluate whether LLM-generated tests can detect real-world breaking changes introduced by dependency updates across library versions.

\section{Threats to Validity}
\label{sec:threats}

One main threat to validity concerns our reliance on statically identified library API usages to guide test generation. Static analysis may miss open-source library calls that occur only at runtime or are obscured by reflection mechanisms. We reduced this threat by generating tests for every library API a client uses in its codebase. More importantly, this study explores a test generation setting to examine whether LLMs can generate tests based on how client code uses library APIs rather than on information about the breaking change itself. However, our results also indicate that providing richer runtime or library code-change information could improve breaking-change detection.

Another threat concerns our use of single-shot prompting without any repair loop. This is a deliberate design choice. We aim to measure the ability of LLMs to write tests that capture breaking changes without any additional engineering, such as iterative repair or tool use. Therefore, our reported detection rates are lower-bound results and could be improved with agentic pipelines or repair-based strategies.

LLM test generation is inherently stochastic, meaning that repeated generations may produce different tests and, consequently, different detection outcomes. We reduce this source of variation by setting the temperature to zero. However, we generate only one response for each focal method to control the cost of generation and execution in the experiment, and therefore do not measure the variability of results across repeated generations. Future work could generate multiple samples per focal method to quantify this variation and assess the extent to which additional samples improve breaking-change detection.

The final threat concerns the transferability of our approach beyond Java/Maven projects. While adapting to other Java testing frameworks requires only minor changes, such as adjusting the prompt, applying our approach to other testing frameworks(e.g PyTest), build systems (e.g., Gradle), languages (e.g., Kotlin), or dependency ecosystems may require additional adaptations. Therefore, our findings may not directly generalize to these settings.

\section{Conclusion}
\label{sec:conclusion}

We presented BreakGuard, an approach that generates migration test suites for a client that depends on an open-source library. It statically analyzes all call sites a client uses from a library and generates tests for each focal method that exercises any of the library's call sites.  We evaluated BreakGuard on three LLMs, namely GPT4o, Qwen3-coder, and GPT-OSS, across three context variants, namely minimal, method, and class, on 89 reproducible BUMP instances. Each context variant includes additional context the LLM can use to write tests that capture any breaking changes caused by a dependency upgrade in the client codebase. Our results show three major findings. First, class context offers the best trade-off for detection, as it provides the full class code as additional context, along with other program- and dependency-upgrade-related context, such as the library version to upgrade and the call site the client uses in its focal methods. Second, we found that richer context improves breaking-change detection. Our best configuration (GPT-4o with class context) detected 30\% breaking changes at an average cost of \$0.90 per BUMP instance. Third, LLM-generated tests predominantly catch runtime errors (e.g., missing classes and methods) and largely miss behavioral breaking changes. Our results establish a lower bound, and future work should evaluate
agentic repair for recoverable failures (such as compilation or test execution errors on the pre-breaking version), and dependency-graph-aware focal-method selection to improve targeting of the correct API method to test to improve breaking change detection.

\section{Data Availability Statement}
The implementation of BreakGuard, code, and detailed setup for running the evaluations are available on GitHub. All execution data and logs, along with a link to the code repository, are available on Zenodo \cite{Replication}.
\bibliographystyle{IEEEtran}
\bibliography{Bibliography}

\newpage

\end{document}